\documentclass[runningheads]{llncs}
\usepackage{graphicx}

\usepackage{tikz}
\usepackage{comment}
\usepackage{amsmath,amssymb} 
\usepackage{color}

\usepackage[accsupp]{axessibility}  

\usepackage{booktabs}

\usepackage{cite}
\usepackage{paralist}
\usepackage{amsfonts}
\usepackage{subcaption}
\usepackage[labelsep=period]{caption}
\usepackage{url}

\newcommand{\ie}{{\it i.e.}}

\begin{document}
\pagestyle{headings}
\mainmatter
\def\ECCVSubNumber{5299}  

\title{Anti-Neuron Watermarking: Protecting  Personal Data Against Unauthorized Neural Networks} 

\titlerunning{Anti-Neuron Watermarking}
%
\author{Zihang Zou\inst{1} \and
Boqing Gong\inst{2} \and
Liqiang Wang\inst{1}}
\authorrunning{Z. Zou et al.}
%
\institute{$^1$University of Central Florida, $\quad^2$Google Research\\
\email{\{Zihang.Zou,Liqiang.Wang\}@ucf.edu}, $\quad$ 
\email{bgong@google.com}}
\maketitle

\begin{abstract}
We study protecting a user's data (images in this work) against a learner's unauthorized use in training neural networks. It is especially challenging when the user's data is only a tiny percentage of the learner's complete training set. We revisit the traditional watermarking under modern deep learning settings to tackle the challenge. We show that when a user watermarks images using a specialized linear color transformation, a neural network classifier will be imprinted with the signature so that a third-party arbitrator can verify the potentially unauthorized usage of the user data by inferring the watermark signature from the neural network. We also discuss what watermarking properties and signature spaces make the arbitrator's verification convincing. To our best knowledge, this work is the first to protect an \emph{individual} user's data ownership from unauthorized use in training neural networks.  
\end{abstract}

\section{Introduction}

Recent advances in machine learning techniques have put personal data at significant risk. For example, in the scandal of ``Cambridge Analytica"~\cite{wiki:CA}, millions of users' data are collected without consent to train machine learning models for political advertising. To protect personal data and privacy, there have been some legislations in place, such as Europe General Data Protection Regulation~\cite{GDPR} (effective in May 2018), California Privacy Act~\cite{CCPA} (effective in January 2021), and China Data Security Law~\cite{CDSL} (effective in July 2021). They often require that personal data should be ``processed lawfully, fairly and in a transparent manner" and can only be used ``adequately, relevantly and limited to what is necessary in relation to the purposes (`data minimisation')"~\cite{GDPR}. However, there is a lack of methods for detecting personal data breaches from machine learning models, which have increasingly become the primary motivation for a violator to break a user's data ownership because the models' efficacy heavily depends on data.

This paper studies personal image protection (PIP) from unauthorized usage in training deep neural networks (DNNs). The need for PIP arises when users expose their images to digital products and cloud services. In the era of big data and deep learning, a critical concern is that DNN learners may violate users' intents by using their data to train DNNs without authorization. It becomes worse when the DNN models consequently leak private user information~\cite{carlini2019secret, fredrikson2015model, nasr2019comprehensive, shokri2017membership}. However, how can ordinary users know whether their images, which could be a tiny portion of the DNN learner's complete training set, have been used to train a DNN model?

Traditionally, PIP aims to prevent a user's images from duplicating, remixing, or exploiting (e.g., for a financial incentive) without the user's consent and relies on digital watermarking~\cite{cox2002digital, kundur1998digital, meng2018design, tirkel1993electronic,yu2001digital, zhong2020automated}. The digital watermarking enables a user to imprint images with unique patterns, such as signatures, logos, or stamps, to track and identify unauthorized \emph{copies} of their pictures. 

However, the rise of data-dependent deep learning poses another need for PIP, namely, protecting a user's images from unauthorized use in training DNNs. Could watermarking still fulfill this need?

One inspiring observation is that some DNNs do ``memorize'' certain training examples~\cite{carlini2019secret, feldman2020neural, fredrikson2015model} in various ways, offering a user an opportunity to watermark their images to make them memorizable by the DNNs. We say this watermarking scheme is ``anti-neuron'' because its objective is to facilitate a third-party arbitrator to verify a DNN's use of a user's images in training and then hold the DNN learner accountable. However, we have to resolve two questions to make this anti-neuron watermarking work in practice. What watermarks make a user's images memorizable by DNNs? How can the third-party arbitrator verify that the user's images were indeed part of a DNN model's training set? 

\begin{figure}
\begin{center}
    \includegraphics[width=.75\columnwidth]{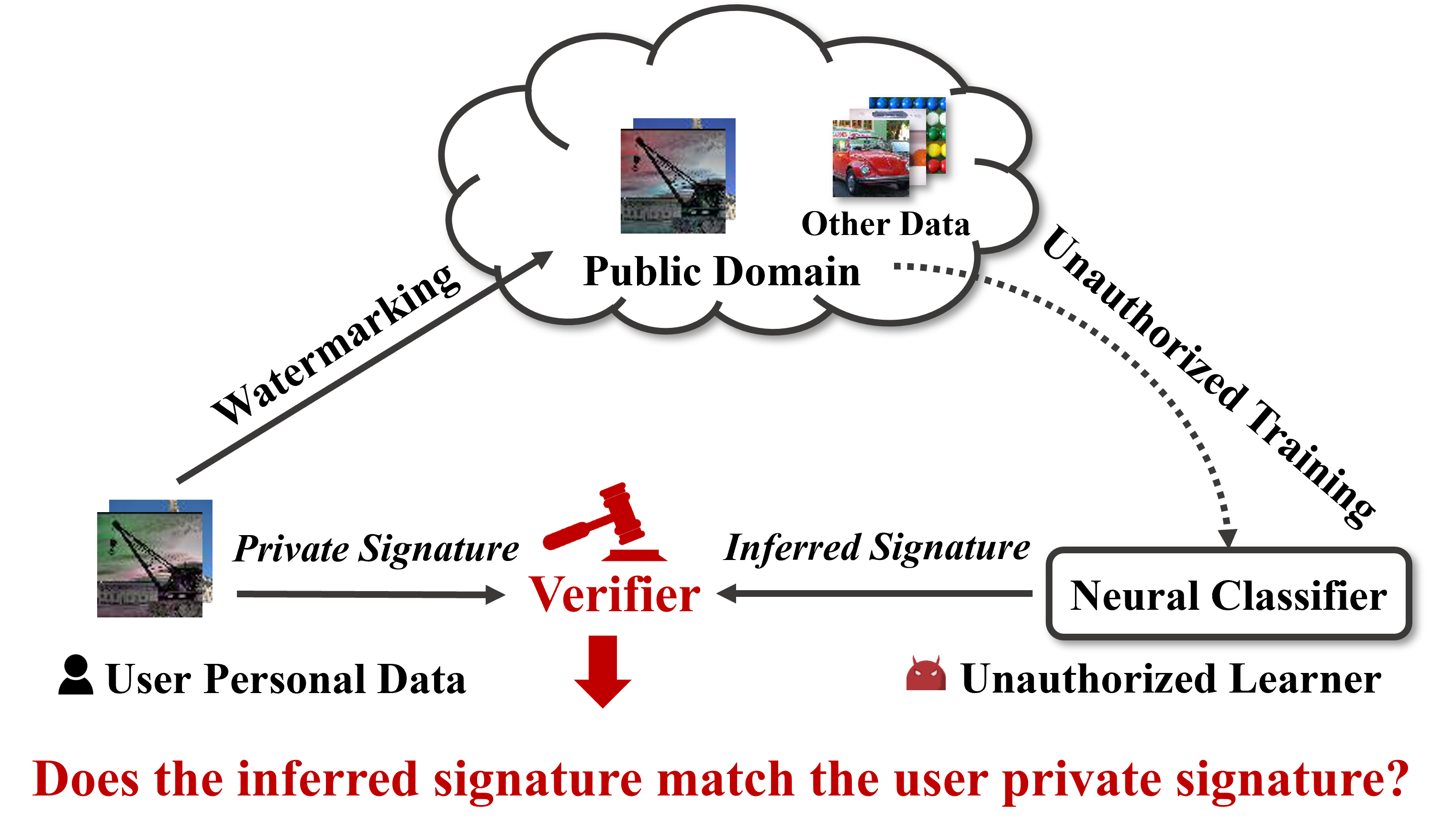}
\end{center}
\caption{Illustration of the anti-neuron watermarking for personal image protection (PIP) against unauthorized neural learners. 
}
\label{fig:illustration}
\end{figure}

To answer the above questions, we first use Figure~\ref{fig:illustration} to formalize the anti-neuron watermarking for PIP against unauthorized DNN learners. First, a user watermarks images using a private signature before sharing them with the public (e.g., social media). An unauthorized learner then collects the user's watermarked images, along with images from  other sources, to construct a training set to train a DNN image classifier. Finally, the user turns to a third-party arbitrator to check whether their images were used to train the DNN model. The arbitrator tries to recover the user's private signature for watermarking from the DNN model and the user's original images with no watermark --- crucially, the arbitrator does not use the user's private signature to recover it. The arbitrator concludes that the user's images were part of the DNN's training set if  the user's private signature can be recovered without knowing it in advance.  

This paper proposes an empirically effective approach to the anti-neuron watermarking, and we leave more rigorous analyses to future work. In particular, a linear color transformation (LCT) in the hue space can be effectively used as a watermarking method. The resultant images remain as appealing as the original ones visually, so  the unauthorized learner would not detect this type of watermark. Moreover, the LCT method is resilient to standard image augmentation techniques used in training neural models. Finally, we show that a DNN classifier indeed tends to memorize the LCT watermarking using extensive experiments. The arbitrator's verification method is simply iterating over the signature space, watermarking the user's original images using each signature, and returning the signature that reaches the lowest DNN classification loss. 

In summary, our main contribution is to formalize the  problem of user-focused anti-neuron watermarking for personal image protection from unauthorized usage in training DNNs. Moreover, we propose the LCT watermarking for ordinary users and a straightforward verification method for the third-party arbitrator, demonstrating a successful anti-neuron watermarking scenario for PIP. Additionally, we raise some critical questions for furthering the study of PIP against DNN learners: 1) What types of watermarking can imprint DNNs the best, especially when a user's watermarked images are only a tiny part of the training set? 2) What makes the imprinting of DNNs possible? Is it the DNN's memorization of training examples? 3) How can a trustworthy arbitrator recover a user's private watermark from DNN and the user's unwatermarked images? 4) How can anti-neuron watermarking work for multiple users? To the best of our knowledge, this work is the first to protect an individual user's data ownership from unauthorized use in training neural networks.  

\section{Related Work}
\textbf{Watermarking} is a long-standing technique to declare ownership of objects. It can be traced back to paper marking~\cite{watermark1998history} at 1282 in Italy, where a watermark was created via changing the thickness of the paper. Digital watermark is later introduced by \cite{tirkel1993electronic} to code an undetectable digital watermark on gray scale images. Yu et al.~\cite{yu2001digital} train a neural network for watermarking to embed hidden information in images. El et al.~\cite{el2006video} add digital watermarks to video frames with neural networks. Zhong et al.~\cite{zhong2020automated} propose an automated and robust image watermarking based on deep neural networks. Recently, watermarking is used to protect the intellectual property of machine learning models~\cite{guo2018watermarking, nagai2018digital, rouhani2019deepsigns, zhang2020model}. These techniques follow similar ideas as trojan attacks~\cite{liu2017trojaning} or backdoor attack~\cite{gu2019badnets}, where models are trained with constructed samples to learn objective behaviors. 

The most relevant works to this paper are {\it dataset tracing} and {\it membership inference}. {\bf Dataset tracing}~\cite{sablayrolles2020radioactive, li2020open, maini2021dataset} protects the intellectual property of a dataset by appending traceable watermarks on data samples. Sablayrolles et al.~\cite{sablayrolles2020radioactive} use pretrained model on a dataset itself to generate ``radioactive data'' to carry the class-specific watermarking vectors in high dimensional feature space. If a learner uses the ``radioactive dataset'' for training, the model's classifier would become more aligned with the watermarking vectors and thus can be used as an evidence of unauthorized usage. This kind of watermarking requires a pretrained model trained with the whole dataset. However, as such prior knowledge about the entire training data is unavailable for a common user,  this kind of technique would be less applicable in real PIP scenarios. 

\textbf{Membership inference} determines if a certain sample is inside a target dataset. Inference attack was first proposed for the attack and defense on medical datasets where users' medical records are extremely sensitive. By comparing genomic data with the statistical information of the training dataset, the presence of certain users can be inferred by attackers~\cite{homer2008resolving}. Shokri et al.~\cite{shokri2017membership} later introduce membership inference attack (MIA) into machine learning models. MIA trains a binary classifier to predict membership, on top of several shadow models being trained with the same data distribution as training. Alternatively, Yeom et al.~\cite{yeom2018privacy} use the average of training error as the threshold to perform MIAs. Sablayrolles et al.~\cite{sablayrolles2019white} improve this threshold with Bayes optimal classifier to search for the best threshold using samples from both training and testing. 

As MIAs determine whether given data samples belong to a training set or not, it is tempting to perform MIAs for personal data protection. However, similar to dataset tracing, as the training data distribution is unknown for common users, neither shadow models~\cite{shokri2017membership} nor threshold~\cite{yeom2018privacy, sablayrolles2019white} can be obtained and existing MIA methods fail to work in protecting personal data.

\section{Problem Statement}
Consider image classification as a case study without loss of generality. Denote by $\mathcal{D}_u$ the set of personal images owned by a common user $u$. Assume $\mathcal{D}_u$ is unique and distinguished among identifiable users, as defined in GDPR~\cite{GDPR}. Suppose the user $u$ plans to expose $\mathcal{D}_u$ online, {\it e.g.}, by sharing them on social media. For the purpose of avoiding potential breach of personal data proprietary, the user watermarks images with a secret signature $k^*$ before sharing them on social media. Denote by $\mathcal{D}_u^*$ the set of watermarked images carrying the signature $k^*$.

An unauthorized learner may use the user's data $\mathcal{D}_u^*$, along with many others', to construct a training set $\mathcal{D}$ to train a DNN classifier $f$ without acquiring the user's permission. It is reasonable to assume that the user's data $\mathcal{D}_u^*$ is only a small portion of the whole training set $\mathcal{D}$ and the user $u$ does not have any prior knowledge about the other users' data.

Let $g\in\mathcal{G}$ denote a watermarking method and $\mathcal{V}$ be a neutral third-party verification method that infers the user's private signature for watermarking without knowing it before. The arbitrator determines whether the user's personal images have been used in the training of neural classifier $f$  as follows,
\begin{gather}
        \mathcal{V}(f, \mathcal{D}_u, \mathcal{G}) = k^* \textbf{ iff } \mathcal{D}_u^* \subseteq \mathcal{D}
\end{gather}
where the user's watermarked images $\mathcal{D}_u^* = g(D_u, k^*), g\in\mathcal{G}$. Namely, if the arbitrator can recover the user's private signature, she/he concludes that the user's images were part of the training set for learning the classifier $f$. 

\section{Approach}
\subsection{The Anti-Neuron Watermarking Method}\label{sec:watermarking}
Recent studies show that DNNs can ``memorize'' some training examples in various ways~\cite{carlini2019secret, feldman2020neural, fredrikson2015model}, and one can recover certain meaningful low-resolution images from DNNs~\cite{fredrikson2015model}. Hence, it is tempting to conduct verification by recovering the user $u$'s images from the neural classifier $f$. However, there are many challenges with this approach. First of all, the model $f$ may memorize some training images but not this user $u$'s. Moreover, even if the model happens to memorize some of this user's images, the recovery success rate is likely low. Existing methods (e.g., \cite{fredrikson2015model})  can recover semantically meaningful images from some DNNs, but they do not resemble any exact training images, to the best of our knowledge. Finally but not the least, the method in \cite{fredrikson2015model} incurs high computation cost, often by many iterations of gradient descent, and assumes that the DNN classifier $f$ is a white box, disclosing its architecture and parameters.

An alternative approach to leveraging DNNs' memorization capability is to check a DNN's loss over a set of training images. Arguably, if the DNN model has memorized a majority of this set of images, the loss should be low. Following the above reasoning, we let a user $u$ watermark her/his images $\mathcal{D}_u$ using a private signature $k^*$ so the user has full control and knowledge of her/his watermarked images $\mathcal{D}_u^*$. This watermarking method eases the third-party arbitrator's job; instead of trying to recover the exact training images, the arbitrator can now search for the watermarking signature that leads to the lowest DNN loss, if the DNN model has memorized many images in $\mathcal{D}_u^*$ watermarked by user $u$.

\subsubsection{Properties for Effective Watermarking.}\label{sec:prop_watermark_function}
Formally, a user $u$ chooses an anti-neuron watermarking function $g\in \mathcal{G}$ and generates the watermarked images as
\begin{equation}
    \mathcal{D}_u^*= \{ g(I, k^*), \forall I\in \mathcal{D}_u \}
    \label{eq:watermarking}
\end{equation}
We discuss the necessary properties needed to make a good anti-neuron watermarking function.
The key is to make the watermarked images, and hence the signature, memorized by DNNs. First, the watermarking function $g$ should \emph{preserve an image's original content}. For example, for a user portrait or selfie, $g$ should not change its identity. Besides, the watermarking function $g$ should be \emph{resilient to common image augmentations} used to train DNNs. The private signature should survive after the learner applies common image augmentations. Furthermore, the space $K$ of watermarking signatures should be \emph{large} and preferably \emph{bounded}, such that the probability of an innocent classifier coincidentally matching the user signature is low, while the signature can be inferred efficiently during verification.

\subsubsection{Linear Color Transformation.}\label{sec:linear_color_transformation} Based on the discussion above, we propose Linear Color Transformation (LCT) as our anti-neuron watermarking. Color provides a large signature space for images. Our watermarking function exploits hue transformation and uses the hue adjustment of images as a signature. Thanks to the sufficiently big hue space, the user's randomly chosen signature is likely different from other users' signatures. Moreover, the randomly chosen signature lifts the user images to a low-density region, making the resultant images be easily memorized by DNNs --- according to Feldman's studies on memorization~\cite{feldman2020neural} and our experiments in Section \ref{sec:memorization}, DNNs tend to memorize images of low-density regions.

Concretely, we first convert the RGB color space into the YIQ color space~\cite{wiki:YIQ} by the following matrix:
\begin{equation}\label{eq:hue}
T_{\text{YIQ}} = 
\begin{bmatrix}
0.299 & 0.587 & 0.114 \\
0.596 & -0.275 & -0.321 \\
0.212 & -0.523 & 0.311 \\
\end{bmatrix}
\end{equation}

In the YIQ color space, hue is represented by two dimensional coordinates, forming a chromaticity diagram. As a result, watermarking images with signature $k$ will be conducted by rotating the hue at an angle $\theta_k$ with the following matrix,
\begin{equation}\label{eq:transformation_matrix}
T_{k} = 
\begin{bmatrix}
1 & 0 & 0 \\
0 & cos(\theta_k) & -sin(\theta_k) \\
0 & sin(\theta_k) & cos(\theta_k) \\
\end{bmatrix}
\end{equation}
where $\theta_k=\frac{k\pi}{180}$. Hence, for every pixel $v=[v_r, v_g, v_b]^\intercal$ in image $I$, we can watermark $v$ with signature $k$ by:
\begin{equation}\label{eq:t}
v{'} = g_k v
\end{equation}
where $g_k=T_{\text{YIQ}}\cdot T_k\cdot T^{-1}_{\text{YIQ}}$. 

\paragraph{Making LCT more versatile.} An immediate extension to LCT is to make the color transformation matrix $T_{\text{YIQ}}$ specifiable by users. A user chosen color transformation $T_u$ can further enrich the watermarking signature space. We leave this extension to future work.

\subsection{The Verification Method}\label{sec:verificaiton}

We let a third-party arbitrator independent of the user and DNN learner determine whether the user's images were part of the DNN training set. The arbitrator has to infer a signature from a suspicious DNN classifier $f$ and the user's original, unwatermarked images $\mathcal{D}_u$ without using the private watermark signature $k^*$. If the inferred signature matches the user's private one, we say that the DNN classifier is highly likely trained using the user's images $\mathcal{D}_u^*$. 

Assume that the watermarking function $g(I, k^*)$ does not change the image's class label. Let $y$ denote the class label of image $I\in\mathcal{D}_u$. We design a simple yet effective approach to recovering the watermarking signature:
\begin{equation}
    \hat{k}\leftarrow \arg\min_{k\in \mathcal{K}}\sum_{(I,y)\in\mathcal{D}_u} \mathcal{L}(f(g(I, k)),y)
    \label{eq:recover}
\end{equation}
where $\mathcal{L}$ is a loss (e.g., cross-entropy) for learning the DNN classifier $f$, and $\mathcal{K}$ is the collection of all possible signatures. 

If the inferred signature matches the user's private one, $\hat{k}\approx k^*$, the arbitrator concludes that the DNN learner has used the user's images $D_u^*=\{g(I, k^*),\forall I\in\mathcal{D}_u\}$ as part of the training set for DNN $f$. Otherwise, the DNN learner is likely innocent. 


\subsubsection{The Signature Space $\mathcal{K}$.}\label{sec:signature_space} 
It is important to discuss the success rate of the above verification method. Apparently, the signature space $\mathcal{K}$ should be sufficiently large to reduce the probability of an innocent classifier coincidentally matching a user's watermarked images. For the analysis purpose, we discretize the bounded signature space $\mathcal{K}$  into $N$ equal-sized, non-overlapped slots, each with an interval $2\tau$. We say the recovered signature $\hat{k}$ matches the private one $k^*$ when $\vert \hat{k} - k^* \vert < \tau$. Reserving one slot for no watermarking, the number of valid watermarking signatures is $N-1$. Clearly, the larger $N$ is, the more convincing the verification. 

Some readers might wonder what if there is a large number of users. For example, given 1 million users but a small number $N$ of signatures, would this setting fail the proposed anti-neuron watermarking? The answer is a pleasant no because, importantly, two users could choose the same private watermarking signature as long as their personal images are different, though the chance of using the same signature is low because each user independently chooses a signature. What happens when a user chooses not to watermark her/his images? A well-trained neural classifier should generalize well under the training distribution. Hence, if most training images are not watermarked, given the user’s original unwatermarked data, the recovered signature from the well-trained classifier would approach no watermarking.

It is not necessary to have enormous $N$ to avoid users having duplicated private signatures based on the above discussion. However, a sufficiently large $N$ is still preferred for another reason, DNNs' memorization. Only when $N$ is big, the chance becomes high for a user to  watermark her/his images into a low-density region and hence can be memorized by DNNs. 

A large signature space also benefits the memorization of user signatures. According to the study \cite{feldman2020neural} on memorization, deep neural classifier must memorize atypical examples to perform well on the less frequent examples during inference. Since watermarking shift data distribution via signature from a large space, watermarking is highly likely to lift user images into lower density region and thus being better memorized by neural models.

\subsubsection{Optimization Method and Computational Cost for Signature Inference.} To solve eq.~(\ref{eq:recover}) efficiently, we propose two optimization methods. \textbf{(i) Grid search :} the arbitrator can enumerate all signatures for watermarking and perform grid search over the bounded signature space with a linear computational cost as $O(N)$. If the signature is well memorized by a DNN, the DNN loss will reach minimum when the signature being evaluated equals or closely approximates the private signature used by the user. 
\textbf{(ii) Gradient search:} when the model is accessible, the arbitrator can watermark clean images with a random initial watermark signature and then infer the user's signature by descending along the gradient of training loss with respect to the signature. This technique infers the watermark signature more precisely than grid search and the computational cost might be less for a large $N$. 

\section{Experiments}\label{sec:experiments}

\subsection{Setup}

We evaluate the proposed watermarking in image classification on the Cifar-10 /\ Cifar-100~\cite{cifar}, CUB birds~\cite{WelinderEtal2010} and Tiny ImageNet~\cite{TinyImageNet} datasets.

\noindent\textbf{A User Watermarks Their Personal Data.} A portion of randomly chosen images from a training set (by default, $1\%$ for Cifar and $0.1\%$ for Tiny ImageNet) is defined as a user' personal data. The user data could contain samples from any class. Each user image is watermarked using eq.~(\ref{eq:t}) by a given signature in the space of $[30,60,...,330]$, followed by clipping pixel values to the valid range of [0,1]. By default, we use 60 (\ie, rotating hue by 60 degree) as the signature.

\noindent\textbf{A Learner Trains Neural Classifiers Using Unauthorized User Data.} An unauthorized learner trains neural classifiers using the above watermarked user data along with other training data. Images are randomly cropped, horizontal flipped, and normalized following the common data augmentation practice~\cite{he2016deep, alexnet}. We use ResNet50~\cite{he2016deep} as the default neural classifier and train every model from scratch for 90 epochs. The initial learning rate is $0.1$ and decays by $0.1$ for every 30 epochs. 

\noindent\textbf{A Verifier Infers The Watermark Signature.} Given suspicious neural network models, a third-party verifier infers the user's signature following two approaches discussed above. For \emph{grid search}, we iterate over all candidate  signatures generated by dividing the whole signature space into $N=12$  intervals whose length is $2\tau=2\times15$. For \emph{gradient search}, we exploit gradient descent to learn the signature. To avoid local optimum, multiple initial values are used and the best signature that leads to the lowest loss is returned.

\subsection{Analyzing Effectiveness of Watermarking}
We first show empirically how signatures are memorized by the neural classifiers. Here, we consider a single user  watermarking their data for simplicity. 
(See Appendix for other experiments and the gradient search results.)

\begin{figure}[t]
  \begin{center}
  \begin{subfigure}[b]{0.3\textwidth}
      \includegraphics[width=\textwidth]{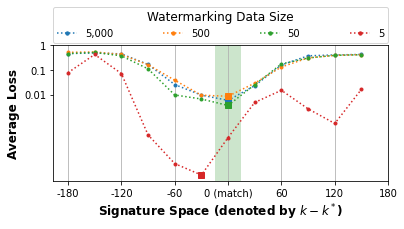}
      \caption{Cifar 10.}
      \label{fig:cifar10_diff_quantity}
  \end{subfigure}
  \begin{subfigure}[b]{0.3\textwidth}
      \includegraphics[width=\textwidth]{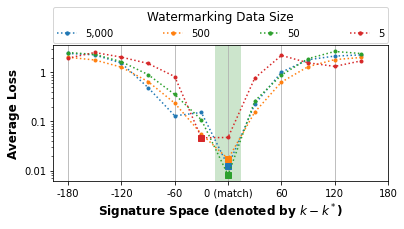}
      \caption{Cifar 100.}
      \label{fig:cifar100_diff_quantity}
  \end{subfigure}
  \begin{subfigure}[b]{0.3\textwidth}
      \includegraphics[width=\textwidth]{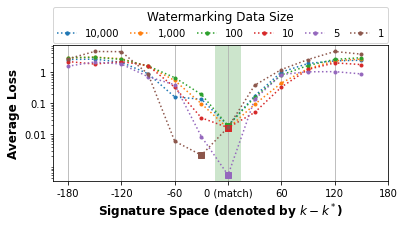}
      \caption{Tiny ImageNet.}
      \label{fig:imagenet_diff_quantity}
  \end{subfigure}
  
  \begin{subfigure}[b]{0.3\textwidth}
      \includegraphics[width=\textwidth]{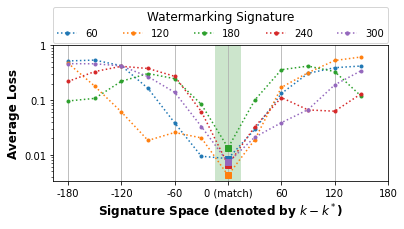}
      \caption{Cifar 10.}
  \end{subfigure}
  \begin{subfigure}[b]{0.3\textwidth}
      \includegraphics[width=\textwidth]{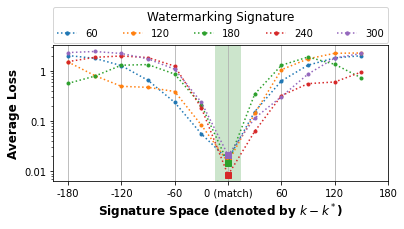}
      \caption{Cifar 100.}
  \end{subfigure}
  \begin{subfigure}[b]{0.3\textwidth}
      \includegraphics[width=\textwidth]{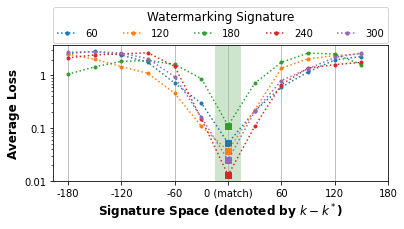}
      \caption{Tiny ImageNet.}
      \label{fig:diff_signatures}
  \end{subfigure}
  
  \begin{subfigure}[b]{0.3\textwidth}
      \includegraphics[width=\textwidth]{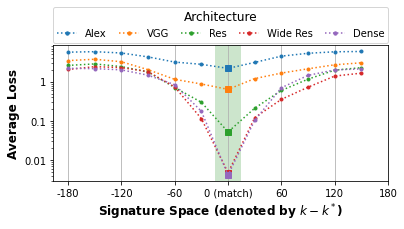}
      \caption{Different architectures.}
      \label{fig:diff_arch}
  \end{subfigure}
    \begin{subfigure}[b]{0.3\textwidth}
      \includegraphics[width=\textwidth]{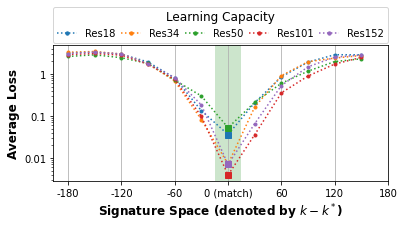}
      \caption{Different capacity.}
      \label{fig:diff_capacity}
  \end{subfigure}
  \begin{subfigure}[b]{0.3\textwidth}
      \includegraphics[width=\textwidth]{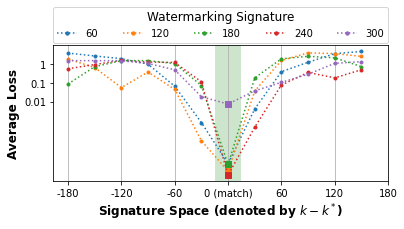}
      \caption{High resolution.}
      \label{fig:high_resolution}
  \end{subfigure}
  \end{center}
    \caption{The first row shows variant of loss for models trained with \textbf{different quantity} of watermarked samples on Cifar and Tiny ImageNet. The second row shows variant of loss for models with \textbf{different signatures} on Cifar and Tiny ImageNet. The third row shows results for \textbf{different architectures}, \textbf{different capacity} on Tiny ImageNet and \textbf{high resolution} on CUB-200-Birds. The x-axis represents signature space (denoted by distance to the user private signature, \ie, $k-k^*$), and the y-axis represents the average loss of user data with respect to signature $k$. The green region represents the range for a match ($<2\tau$). If the inferred signature $\hat{k}$ (marked as \textbf{Square marker} indicating the point with minimum loss) lies in the green region ($\vert \hat{k}-k^* \vert < \tau$), it would be a match. Otherwise, it would be a miss.}
    \label{fig:results}
\end{figure}

\begin{enumerate}
\setlength\itemsep{0em}
\item \textit{Different Numbers of Watermarked Samples.} We study how many images are desired for making anti-neuron watermarking successful. The grid search result is shown in Figure~\ref{fig:results} (a, b, c) using eq.~(\ref{eq:recover}). It is visually clear that most of the models achieve the minimum loss near the watermark signature, within the range of matching $\vert \hat{k} - k^* \vert<\tau$. However, with less watermarked data ({\it e.g.}, less than 5 samples), the inferred signature with minimum loss does not match the user's private signature. 

\item \textit{Different Watermark Signatures.}\label{sec:diff_watermark_val} We verify whether different watermark signatures work equivalently. We experiment with different signatures on one user's data and show the grid search results in Figure~\ref{fig:results} (d, e, f) for different datasets. From these figures, we observe that all inferred signatures (marked in square) match the user's signatures for watermarking, indicating that different hue adjustments can all be used for anti-neuron watermarking.

\item \textit{Different Neural Classifier Architectures.} We also evaluate the proposed watermarking for different neural classifier architectures, including Alexnet~\cite{alexnet}, VGG~\cite{simonyan2014very}, ResNet~\cite{he2016deep}, Wide ResNet~\cite{zagoruyko2016wide} and DenseNet~\cite{huang2017densely} trained with default settings. As shown in Figure~\ref{fig:diff_arch}, all inferred signatures match the user's, implying that our watermarking approach works well against a large variety of deep neural networks. 

\item \textit{Different Learning Capacities of Models.}
We further investigate whether a model memorizes watermark signatures better when the model has more learning capacity (e.g., more parameters, deeper or wider) by exploring the ResNet family. As shown in Figure~\ref{fig:diff_capacity}, as the networks go larger and deeper, the loss decreases faster and reaches the minimum around the watermark signature more sharply.

\item \textit{High Resolution Images.} In Figure~\ref{fig:high_resolution}, we present our result on CUB-200-Birds, a fine-grained dataset with high-resolution images of $448\times 448$. We use pretrained ResNet50 from ImageNet and conduct a transfer learning on CUB-200-Birds. The dataset has fewer than 6000 images for training, and we assume the user has 60 images (1\%) for watermarking. Strong data augmentations~\cite{yu2018deep} are used to boost performance, including color jitter, random crop, random resize, random scale and random horizontal flip. Even under the strong data augmentations and the transfer learning setting, the result shows that ResNet50  memorizes the user's signature well. 
\end{enumerate}

\subsection{Analyzing Properties of Watermarking}
We then evaluate how the properties discussed in Section~\ref{sec:prop_watermark_function} would help anti-neuron watermarking. 

\begin{figure}
  \begin{center}
  \small
  \begin{subfigure}[b]{0.24\textwidth}
      \includegraphics[width=\textwidth]{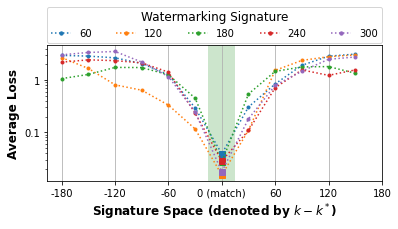}
      \caption{\tiny Cut Out.}
      \label{fig:aug_co}
  \end{subfigure}
  \begin{subfigure}[b]{0.24\textwidth}
      \includegraphics[width=\textwidth]{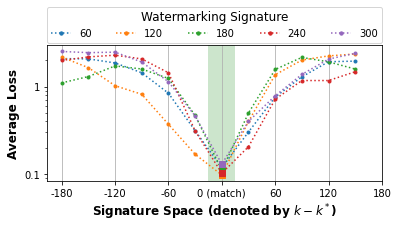}
      \caption{\tiny Label Smoothing.}
      \label{fig:aug_ls}
  \end{subfigure}
  \begin{subfigure}[b]{0.24\textwidth}
      \includegraphics[width=\textwidth]{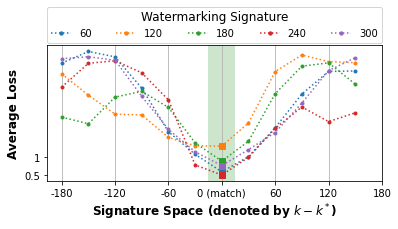}
      \caption{\tiny Gaussian Noise.}
      \label{fig:aug_gn}
  \end{subfigure}
  \begin{subfigure}[b]{0.24\textwidth}
      \includegraphics[width=\textwidth]{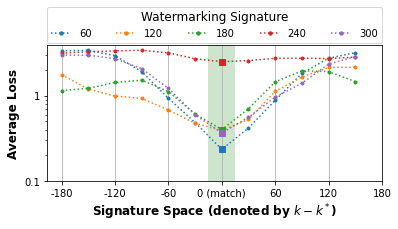}
      \caption{\tiny Adversarial Training.}
      \label{fig:aug_adv}
  \end{subfigure}
  \begin{subfigure}[b]{0.24\textwidth}
      \includegraphics[width=\textwidth]{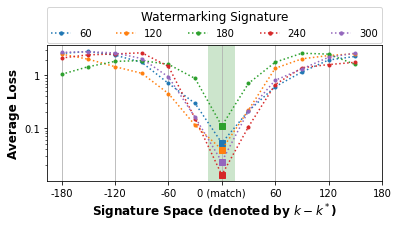}
      \caption{\tiny Differential Privacy.}
      \label{fig:aug_dp}
  \end{subfigure}
  \begin{subfigure}[b]{0.24\textwidth}
      \includegraphics[width=\textwidth]{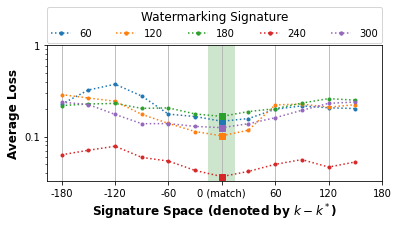}
      \caption{\tiny Color Jitter.}
      \label{fig:aug_cj}
  \end{subfigure}
  \begin{subfigure}[b]{0.24\textwidth}
      \includegraphics[width=\textwidth]{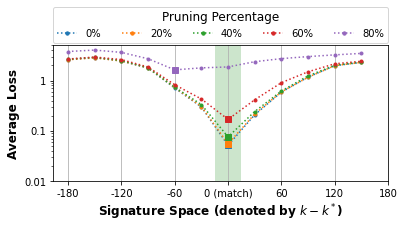}
      \caption{\tiny Pruning.}
      \label{fig:aug_prune}
  \end{subfigure}
  \begin{subfigure}[b]{0.24\textwidth}
      \includegraphics[width=\textwidth]{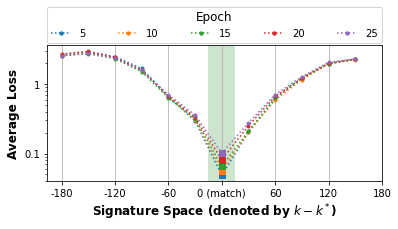}
      \caption{\tiny Fine-tuning.}
      \label{fig:aug_fine_tuning}
  \end{subfigure}
  \end{center}
    \caption{The variation of model loss for \textbf{different data augmentations}. Only the color jitter can significantly narrow down the loss difference between signatures.}
    \label{fig:resilent_to_data_aug}
\end{figure}

\noindent \textbf{Resilience to Data Augmentation.}
We evaluate if our anti-neuron watermarking is resilient against various common data augmentations, especially those involving random hue transformations. We apply random crop and random horizontal flip in all our experiments following~\cite{alexnet,he2016deep}. Besides, several widely adopted data augmentations including random cutout~\cite{devries2017improved}, label smoothing~\cite{szegedy2016rethinking}, Gaussian noise~\cite{cohen2019certified}, adversarial training~\cite{madry2018towards} and differential privacy~\cite{dwork2008differential} are evaluated. Finally, we test color jittering~\cite{alexnet}, which includes brightness, saturation, contrast and \textit{the same} hue transformation we used for watermarking. As shown in Figure~\ref{fig:resilent_to_data_aug}, the watermark signatures can be inferred correctly for the aforementioned data augmentations. This shows empirically that LCT is an effective anti-neuron watermarking approach because it is resilient to common data augmentations in neural networks’ training. We also evaluate privacy preserving techniques such as differential privacy~\cite{dwork2008differential}. Since we infer the signature using all user images, noise added to the output would be reduced by taking an average.
Beside, we also consider two common \emph{defense techniques} against watermarking: pruning and fine-tuning. We follow common settings~\cite{zhang2020passport} and find in Figure~\ref{fig:aug_prune} and Figure~\ref{fig:aug_fine_tuning} that LCT is also resilient to these defense methods.

\begin{figure}
  \begin{center}
    \begin{subfigure}[b]{0.1\textwidth}
      \includegraphics[width=\textwidth]{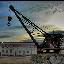}
      \caption{\tiny Clean}
    \end{subfigure}
    \begin{subfigure}[b]{0.1\textwidth}
      \includegraphics[width=\textwidth]{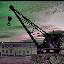}
      \caption{\tiny Hue}
    \end{subfigure}
    \begin{subfigure}[b]{0.1\textwidth}
      \includegraphics[width=\textwidth]{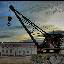}
      \caption{\tiny 0.1}
    \end{subfigure}
    \begin{subfigure}[b]{0.1\textwidth}
      \includegraphics[width=\textwidth]{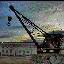}
      \caption{\tiny 0.3}
    \end{subfigure}
    \begin{subfigure}[b]{0.1\textwidth}
      \includegraphics[width=\textwidth]{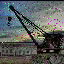}
      \caption{\tiny 0.5}
    \end{subfigure}
  \end{center}
    \caption{Illustration of clean, hue-transformed and less noticeable watermarked samples. 0.1,0.3 and 0.5 are the intensity of blue channel for selected pixels.}
  \label{fig:unnoticeable}
\end{figure}

\noindent \textbf{Less Noticeable Watermarking.}\label{sec:unnoticeable} In Section~\ref{sec:prop_watermark_function}, we discuss the watermarking should not change the major content of an image, and one of the desired features is to make watermarking unnoticeable to human. Rather than changing the hue of images globally, we study an alternative technique from traditional watermarking proposed in~\cite{yu2001digital}. It adjusts the blue channel's intensity on pre-selected pixels. In this work, we adjust the intensity as a watermark signature on 512  randomly selected pixels. As shown in Figure~\ref{fig:unnoticeable}, the general appearance of watermarked images are less noticeable than changing the hue globally (adjusting hue for 4096 pixels by 60). For $\tau=0.1$ and watermark signatures (the blue channel’s intensity), 0.1, 0.3, 0.5,  0.12, 0.28, 0.44 are used in inference on selected pixels, matching the user's watermark signature. However, this kind of watermarking has its limitation. It introduces noise to images, and the images could look noisy when the color themes are dominated by red or green. To solve this problem, we may find some transformations that are invisible to humans but easy to learn by neural classifiers. We leave this challenge to future work.

\subsection{Analyzing Signature Space for Verification}
In Section~\ref{sec:signature_space}, we discuss how the signature space could affect watermarking from the perspective of a third-party verifier. Here, we show experimentally how to have a trustful signature space for a convincing verification. 

\noindent \textbf{When User Data Was Not Used to Train Neural Classifiers.} From previous experiments, we show that the inferred signature  matches a user's private watermark signature if the user's watermarked data have been used in training. Here we show the inferred signature approaches no watermarking when the user data was not used for training. To this end, we construct held-out users using auxiliary unseen data from validation. Pretrained models from Figure~\ref{fig:diff_signatures} are used to infer signature from the held-out users. Not surprisingly, the inferred signatures approach 0 (no watermarking) for the held-out users, with $\vert \hat{k}-0 \vert = 4.3\pm1.4$. 

\noindent \textbf{Multiple Users with User-specific Watermarking Signatures.}\label{subsubsec:MultipleUsers} 
We examine multiple users for several scenarios. The training set of Tiny ImageNet is equally divided into 1,000 users. Then we evaluate the effectiveness of watermarking for different ratio of users exploiting the same LCT (eq.~(\ref{eq:hue})) or different LCTs. For the later, we sample $3\times3$ matrices from a uniform distribution $T_u\sim\mathcal{U}(-1,1)$ per user. Each user chooses a random signature from $[30, 60, ..., 330]$, and $\tau$ is set to 15. 

\begin{figure}
  \begin{center}
  \begin{subfigure}[b]{0.35\textwidth}
      \includegraphics[width=\textwidth]{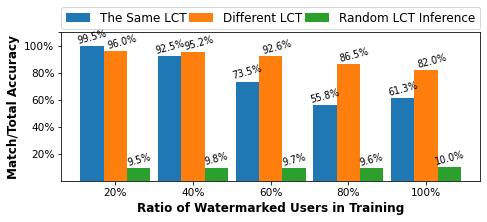}
        \caption{}
      \label{fig:diff_matrix}
  \end{subfigure}
  \begin{subfigure}[b]{0.25\textwidth}
      \includegraphics[width=\textwidth]{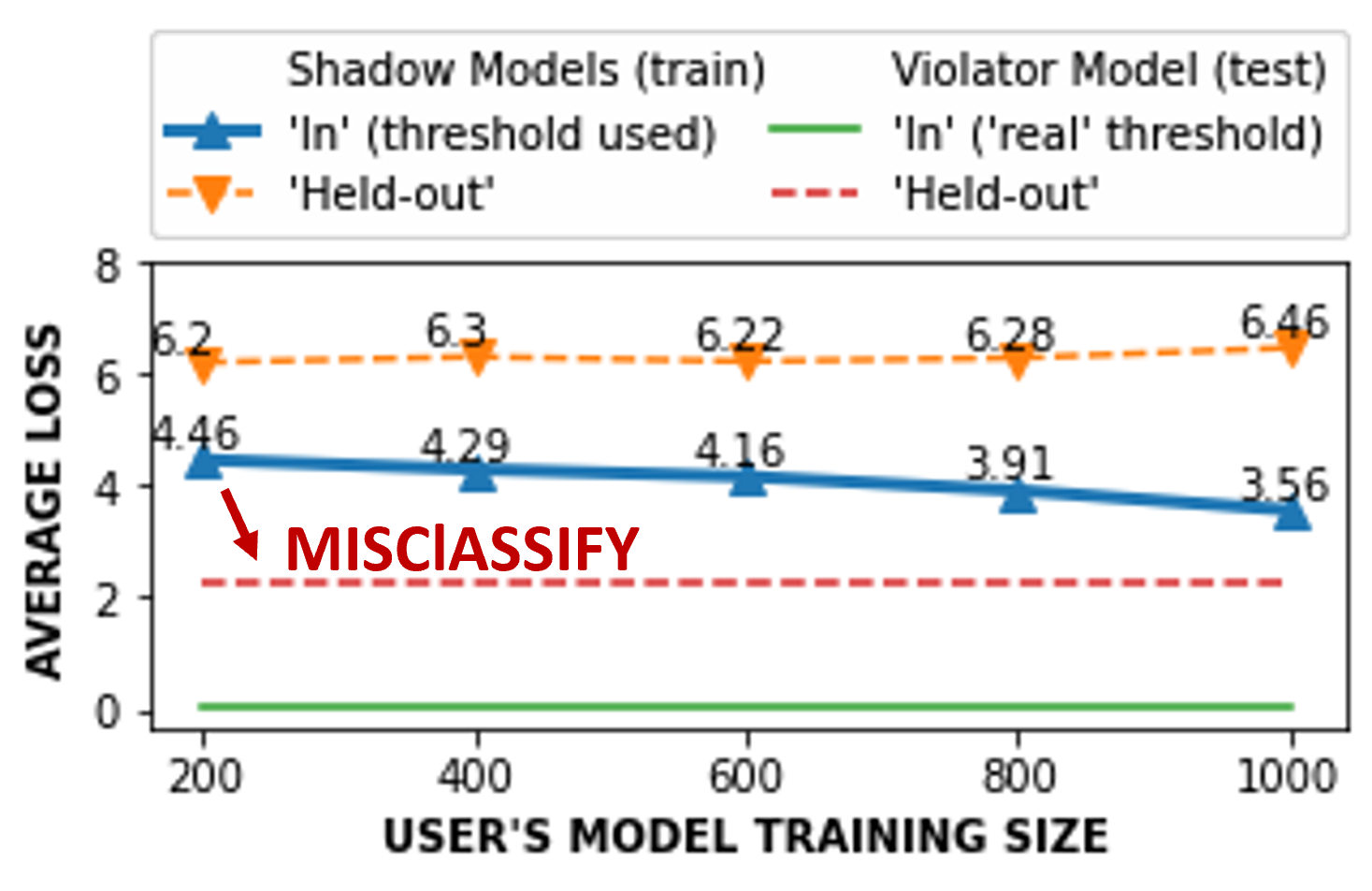}
    \caption{}
      \label{fig:user_level_mia}
  \end{subfigure}
  \begin{subfigure}[b]{0.35\textwidth}
      \includegraphics[width=\textwidth]{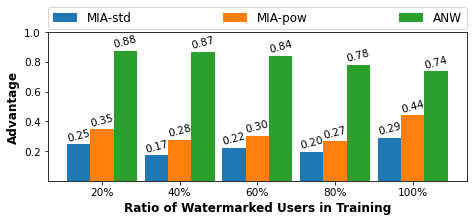}
    \caption{}
      \label{fig:comparison_mid}
  \end{subfigure}
  \end{center}
  \caption{(a) Watermarking performance for \textbf{multiple users}. (b) MIAs fail to work with user-specific shadow models because insufficient training data leads to much higher threshold. (c) Our ANW \textit{vs.} MIAs using the adv\textsuperscript{P} metric.}
 \end{figure}

From the result shown in Figure~\ref{fig:diff_matrix}, we can observe that when $20\%$ of users watermark their images using LCT, their watermark signatures can be inferred correctly for almost all the users. As this ratio increases, the matching accuracy drops significantly if the users use the same LCT. However, if they use different LCTs, the matching accuracy remains above $80\%$ even when all users data are watermarked independently. We also evaluate a special case when an adversary infers signatures using an arbitrary LCT. The arbitrary LCTs ($T_{u^{'}}\sim\mathcal{U}(-1,1)$) only achieve $10\%$ matching accuracy, which is $70\%$ less when LCTs are given. Such results indicate users can use unique and user-specific watermarking for a better protection rate when other users may also exploit watermarking.

\subsection{Comparisons with Related Methods}

\noindent \textbf{Comparing to Membership Inference Attacks.} We compare our anti-neuron watermarking (ANW) with two \emph{threshold-based} membership inference attacks (MIAs): In MIA-std~\cite{yeom2018privacy}, few samples are known in violator's training set and their average training loss would be used as threshold $\epsilon$ to infer membership (if $\mathcal{L}_t<\epsilon$, target sample $t$ is in training, {\it vice versa}.); In MIA-pow~\cite{sablayrolles2019white}, few samples in training and held-out together determine a better threshold. 

There are 3 scenarios making MIAs not applicable for PIP. 
\textbf{(i)} as a user has no knowledge about the violator's training distribution, MIAs cannot be applied as neither shadow models nor threshold can be obtained. 
\textbf{(ii)} Users could train shadow model using their own data and perform MIAs via shadow models~\cite{shokri2017membership}, but MIAs would not work well because shadow models trained with user data would produce a much larger loss than the learner' models. Consequently, if a threshold is chosen from shadow models trained with user data, held-out sample would be misclassified as ``in training'' because the learner's model would produce smaller loss for both ``training'' and ``held-out'' samples. Here we compare MIAs by creating 10 users with individual data. 10 shadow models are trained respectively and MIA-std is performed with the threshold of average training loss. For ANW, 10 users data are watermarked with different signatures. Under different settings for user data size, we find out that ANW achieves 100\% matching accuracy while MIA-std achieves 50\% (misclassify all held-out samples). Figure~\ref{fig:user_level_mia} shows how the threshold from shadow models fails to classify sample membership for the learner. Similarly, MIA-pow along with other MIAs~\cite{shokri2017membership, maini2021dataset} relied on shadow models would also fail in this settings.
\textbf{(iii)} At last, even if a user acquires all necessary information, the MIA results would still be less convincing. As MIAs only provide binary output (True/False), it is difficult to convince the verifier when random guess can still achieve about $50\%$ success rate. To quantifiably and fairly compare with MIAs, we extend the classic \emph{membership advantage}~\cite{yeom2017unintended} into \emph{protection advantage} (adv\textsuperscript{P}) considering both accuracy and fidelity: 
\begin{equation}
     \text{adv\textsuperscript{P}} = \int_x (\mathbb{P}_{e}(y=True|x) - \mathbb{P}_r(y=True|x))dx
\end{equation}
The adv\textsuperscript{P} metric quantifies quality of protection through the expectation gap between the empirical successful inference ($\mathbb{P}_e$) and successful random guess ($\mathbb{P}_r$). Note that the \emph{membership advantage}~\cite{yeom2017unintended} is a special case of adv\textsuperscript{P} metric when the second term is $0.5$. For a Bernoulli experiment, the above formula could be calculated as $\frac{M-Np}{N}$, where $M$ is the total matches in $N$ experiments and $p$ is the probability of a correct random guess. With above metric, we conduct comparison between our watermarking and two threshold-based MIAs~\cite{yeom2018privacy, sablayrolles2019white} (See appendix for experiment settings.). 
As shown in Figure~\ref{fig:comparison_mid}, our ANW significantly outperforms the MIA approaches under the adv\textsuperscript{P} metric with both accurate and convincing inference, showing that watermarking is a feasible method in the PIP problem.

\noindent \textbf{Comparing to Dataset Tracing~\cite{sablayrolles2020radioactive}.} Dataset tracing~\cite{sablayrolles2020radioactive} exploits pretrained classifier to generate traceable data. If neural classifier learns such dataset, the decision boundary of classifier would become more aligned with watermarking vectors (\ie, cosine similarity becomes higher). In Table~\ref{tbl:comparison_with_dataset_tracing}, we compare this approach~\cite{sablayrolles2020radioactive} with ours when only 0.1\%  data being watermarked. 
For dataset tracing, it is computed by the classifier's weight vector and the watermarking vector. And for our method, it is computed by inferred and user watermarking signatures. The experimental results show that it is easier to memorize low dimensional signatures as our watermarking method lifts the cosine similarity significantly after training with tiny portion of watermarked data.

\begin{table}
\begin{center}
\small
\begin{tabular}{l|cc}
\hline
& Before Training & After Training \\
\hline
Dataset Tracing\cite{sablayrolles2020radioactive} & $-0.005\pm0.030$ & $-0.005\pm0.015$ \\
Our Watermarking & $0.045\pm0.52$ & $0.999\pm0.001$ \\
\hline
\end{tabular}
\end{center}
\caption{\textbf{Cosine similarity}. With only $0.1\%$ data watermarked, the dataset tracing shows almost no effects while our watermarking method improves the similarity to almost 1 after training.}
\label{tbl:comparison_with_dataset_tracing}
\end{table}

\subsection{Improving Memorization by Watermarking }\label{sec:memorization}
Finally, we explore empirically why watermarking is effective against neural classifier by revisiting ``Memorization Value Estimate" (MAE)~\cite{feldman2020neural}. MAE measures the generalization gap (difference of prediction between models trained with/without certain data) to quantify the memorization ability of neural networks toward such data. A higher MAE after watermarking indicates the model tends to memorize watermarked data than original data. 
For user data, we observe the MAE increases from $26.8\%$ to $34.8\%$ after applying watermarking, indicating that our approach increases memorization of user data and thus the signature would be easier memorized along with user data.

\section{Conclusion}
In this paper, we introduce a new personal data protection problem against unauthorized neural model training. To protect user personal data, we propose an anti-neuron watermarking approach based on linear color transformation. By watermarking user's images with private signature using LCT, unauthorized usage of user personal data can be verified by a third-party neutral arbitrator. Through extensive experiments, we show empirically that LCT-based watermarking is effective in protecting user data from unauthorized usage in a various realistic settings. 

\noindent\textbf{Acknowledgements:} this work was supported in part by NSF-1704309 and NSF-1952792.

\clearpage
%
%
\bibliographystyle{splncs04}
\bibliography{egbib}

\appendix
\section{Background}
In the following sections, we present the background knowledge of personal data protection problem studied in our paper. We first quote the legislative definition of ``personal data protection'' from GDPR~\cite{GDPR} and then we discuss the scope of our problem comparing to general data protection. Finally, some ``data proprietary'' problem are addressed in personal data protection.

\noindent \textbf{Definitions of Terms from GDPR. } To help readers better understand the terms we used in the manuscript, we quote several term definitions from \cite{GDPR} as following:

``\emph{Personal data} means  any information relating to an identified or identifiable natural person (`data subject')." 



``\emph{Third party} means a natural or legal person, public authority, agency or body other than the data subject, controller, processor and persons who, under the direct authority of the controller or processor, are authorised to process personal data."

``\emph{Consent} of the data subject means any freely given, specific, informed and unambiguous indication of the data subject’s wishes by which he or she, by a statement or by a clear affirmative action, signifies agreement to the processing of personal data relating to him or her."

``\emph{Personal data breach} means a breach of security leading to the accidental or unlawful destruction, loss, alteration, unauthorised disclosure of, or access to, personal data transmitted, stored or otherwise processed."

From above definitions, we show the basic idea of GDPR: a legislation approach to protect users personal data from being utilized by third-party without users' consent. More details of GDPR and similar laws can be found online~\cite{GDPR, CCPA, CDSL}. In our paper, we focus on the technical perspective of personal data protection by exploiting anti-neuron watermarking to verify unauthorized usage of user's personal images.

\noindent \textbf{Personal Data Protection {\it v.s} General Data Protection.}
As our goal is to protect ``Personal Data'' ({\it i.e.} user data shall be related to identifiable natural person~\cite{GDPR}), we assume each user's personal images can be distinguished from the others. This assumption differs ``personal data protection'' from ``general data protection'' problem, where a user's data are not necessarily to be identified from other users'. In ``general data protection'', our anti-neuron watermarking would be less applicable. Because neural models learn both watermarked and unwatermarked data, the inferred signature would be a signature value between the watermarked and unwatermarked. We leave this challenge for future studies.

\noindent \textbf{``Data proprietary'' in Personal Data Protection.} One traditional usage of digital watermarking is to protect data proprietary. To verify proprietary, the space of watermarking signature needs to enormous such that each signature ({\it i.e.}, a hand-writing signature) can be considered as a unique identifier for user data. However, in our problem settings, our protection focuses on verifying unauthorized usage ({\it i.e.},  without ``consent'') of user data (images that user already has ``data proprietary'' on) instead of verifying data proprietary. Since the low-dimensional signature is only used to verify unauthorized usage of images, our signature space does not need to be as large as ``data proprietary'' problem. Readers might be curious what if adversary exploits user's watermarking function and reverts a signature on arbitrary images that leads to minimal loss of arbitrary neural models. Can adversary accuse data privacy breach for these models? Can adversary claim ownership of these arbitrary data? The answers are no. According to our previous analysis on signature space, the signature inferred on arbitrary images is highly likely to be no watermarking, a signature value excluded from valid watermarking signatures. On the other hand, adversary cannot claim data propriety on arbitrary data as they cannot prove that the data are legally ``relating to'' adversarial users. 

\section{Preliminary Study}
\noindent \textbf{Verification by Recovering Watermarking Pattern.} Recent studies show that DNNs can ``memorize'' some training examples in various ways~\cite{carlini2019secret, feldman2020neural, fredrikson2015model}, and one can recover certain meaningful low-resolution images from CNNs~\cite{fredrikson2015model}. Motivated by these studies, we perform a preliminary study on traditional watermarking technique by appending a special pattern ({\it e.g.}, a sticker) on images. We train a ResNet50 on 100 randomly selected user samples with a cat pattern as watermark in Tiny ImageNet. Similar to model inversion~\cite{fredrikson2015model}, we use a learnable Gaussian variable on users' images and minimize the classification loss to reconstruct watermark pattern with this variable. 

\begin{figure}
  \begin{center}
  \begin{subfigure}[b]{0.15\textwidth}
      \centering
      \includegraphics[width=\textwidth]{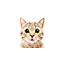}
  \end{subfigure}
  \begin{subfigure}[b]{0.15\textwidth}
      \centering
      \includegraphics[width=\textwidth]{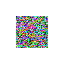}
  \end{subfigure}
  \begin{subfigure}[b]{0.15\textwidth}
      \centering
      \includegraphics[width=\textwidth]{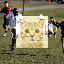}
  \end{subfigure}
  \end{center}
    \caption{Illustration of watermark pattern (left), recovered pattern (middle) and watermarked images (right). Recovering small visible watermark fails with getting noise. }
  \label{fig:traditional_watermark}
\end{figure}

As shown in Figure~\ref{fig:traditional_watermark}, although the reconstructed pattern can achieve $100\%$ accuracy and minimal classification loss, neural models cannot memorize such watermark pattern, as no meaningful pattern can be recovered. We believe that this could be caused by convolution operation where all spatial information of pixels are being ignored during training. This experiment indicates that it is difficult for neural learners to memorize such kinds of watermarks, comparing to the watermarks using color-based transformation. 

\section{Implementation Details}

\noindent \textbf{Datasets.} We evaluate our anti-neuron watermarking on Cifar~\cite{cifar}, Tiny ImageNet~\cite{TinyImageNet} and CUB-200-Birds~\cite{WelinderEtal2010}. Cifar are widely adopted datasets with 50,000 training samples, 10,000 testing samples for 10 and 100 classes, respectively. Tiny ImageNet is a selective subset of ImageNet, containing 100,000 training samples, 10,000 validation samples and 10,000 testing samples for 200 classes. Each sample is $3\times 64 \times 64$. Since testing labels are not publicly available, we report models' validation accuracy for models' performance. CUB-200-Bird is a high resolution $448\times448$ fine-grained dataset containing 200 bird species, with 5994 training samples and 5794 testing samples. 

\noindent \textbf{Watermark User Data.} Each user image is watermarked by given $3\times 3$ LCT function with a signature $k$ followed by a pixel value clipping. A clipping is needed because LCT could cause overflow on some pixels' values. The clipping operation in theory will break the differentiable property of watermarking function and thus hinder gradient based optimization ({\it e.g.,} stochastic gradient decent). However, we find empirically that this operation does not affect signature inference. Hence, we conduct clipping after the LCT in all our experiments.

\noindent \textbf{Data Preparation.} For training, each image is first converted from $[0, 255]$ to $[0, 1]$, and watermarked if it belongs to the user whose images need to be protected. Then data augmentation and normalized are applied to improve training. During signature inference, each image is converted from $[0, 255]$ into $[0, 1]$, then watermarked and normalized. 
The normalization mean and variance of RGB channels are $(0.5,0.5,0.5)$ and $(0.5,0.5,0.5)$ for Cifar, $(0.485, 0.456, 0.406)$ and $(0.229, 0.224, 0.225)$ for Tiny ImageNet and CUB-Birds, respectively.

\noindent \textbf{Grid Search Settings.} We iterate all the possible signatures in our grid search experiments. The signatures are generated by dividing the whole signature space into $N\times 2\tau$ intervals. 

\noindent \textbf{Gradient Search Settings.} During signature inference, we recover signature using unwatermarked user data. These user data are combined into one mini-batch and stochastic gradient decent is used for optimization over signatures. The initial learning rate is 0.1 and decays 0.1 each 100 epochs, with 300 epochs in total. To avoid local minima, we select initial values from all possible signatures and report the signature that lead to minimal loss. Comparing to grid search, this optimization approach is much more computational expensive and thus our results will be mainly on grid search for simplicity.

\noindent \textbf{Source Code.} Our experimental source code will be publicly available (code is available in supplementary folder). All our experiments are implemented by Pytorch. Readers can freely explore our proposed watermarking approach with the code supplied.

\section{Additional Results for Effectiveness of Watermarking}


\noindent \textbf{Gradient Search Results.} In the main manuscript, we present the results for grid search in various settings. Here, we show the additional gradient search results on Tiny ImageNet.

\begin{enumerate}
\setlength\itemsep{0em}

\item \textit{Different sizes of Watermarked Samples.} We first present our result with different sizes of watermarkred samples in Table~\ref{tbl:diff_ratio_resnet50_tiny_imagenet}. Being trained by classifier, user's watermarked images achieve lower loss than the clean images. And watermarking does not affect classification performance for getting similar testing accuracy. These results show that given sufficient data, neural classifier could memorize watermark signature on user's data pretty well. 

\begin{table}
\begin{center}
\small
\begin{tabular}{c|c|c|c|c}
\hline
\shortstack{\# of\\Data} & \shortstack{Model\\Acc} & \shortstack{Watermark \\ Loss} & \shortstack{Clean\\ Loss} & \shortstack{Inferred\\signature} \\
\hline
10,000 & 55.6 & 0.019 & 0.161 & 59.0 \checkmark\\
1,000  & 55.8 & 0.016 & 0.541 & 56.4 \checkmark \\
100  & 54.9 & 0.019 & 0.668 & 59.5 \checkmark \\
10  & 54.5 & 0.017 & 0.329 & 48.9 \checkmark \\
5  & 54.6 & 0.001 & 0.397 & 60.99 \checkmark \\
1  & 55.9 & 0.004 & 0.006 & 17.0 $\times$ \\
\hline
\end{tabular}
\end{center}
\caption{Inferred signatures for models trained with \textbf{different sizes} of watermarked data on Tiny ImageNet. The watermark signature is 60 with $\tau=15$.}
\label{tbl:diff_ratio_resnet50_tiny_imagenet}
\end{table}

\item \textit{Different Watermark Signatures.} We present the gradient search result for different signatures on Tiny ImageNet in Table \ref{tbl:diff_signature}. The result shows that the watermarking for single user does not affect models' training as models' accuracy are similar for different signatures. User watermarked images achieve lower average loss than original images, indicating unauthorized training models can memorize watermarked images in ``some way''. And we achieve minimal loss nearby watermarking signature, bounded by predefined threshold $\tau=15$. This experiment shows that different signatures of watermarking work equivalently.

\begin{table}
\begin{center}
\small
\begin{tabular}{c|c|c|c|c}
\hline
\shortstack{Watermark\\signature} & \shortstack{Model\\Acc} & \shortstack{Watermark\\Loss} & \shortstack{Clean\\Loss} & \shortstack{Inferred\\signature} \\
\hline
60  & 54.9 & 0.019 & 0.668 & 59.5 \checkmark\\
120 & 55.6 & 0.070 & 0.895 & 120.2 \checkmark\\
180 & 53.8 & 0.030 & 1.189 & 178.8 \checkmark\\
240 & 56.3 & 0.025 & 1.118 & 242.1 \checkmark\\
300 & 55.7 & 0.016 & 0.839 & 306.1 \checkmark\\
\hline
\end{tabular}
\end{center}
\caption{Watermarking for \textbf{different signatures} for ResNet50 on Tiny ImageNet.}
\label{tbl:diff_signature}
\end{table}

\item \textit{Different Neural Classifier Architectures.} We present the gradient search result for different architectures on Tiny ImageNet in Table \ref{tbl:diff_arch}. Alexnet~\cite{alexnet}, VGG~\cite{simonyan2014very}, ResNet~\cite{he2016deep}, Wide ResNet~\cite{zagoruyko2016wide} and DenseNet~\cite{huang2017densely} are evaluated in this experiment. The watermark signature is 60 and $\tau=15$.

\begin{table}
\begin{center}
\small
\begin{tabular}{c|c|c|c|c}
\hline
Architecture & \shortstack{Model\\Acc} & \shortstack{Watermark\\Loss} & \shortstack{Clean\\Loss} & \shortstack{Inferred\\signature} \\
\hline
 Alex & 38.0  & 2.189 & 3.175 & 56.9\checkmark \\
 VGG & 57.3 & 0.640 & 1.151 & 52.8\checkmark \\
 Res & 54.9 & 0.019 & 0.668 & 59.5\checkmark \\
 Wide Res & 56.6 & 0.005 & 0.738  & 58.4\checkmark \\
 Dense & 61.5 & 0.114 & 0.838 & 58.8\checkmark \\
\hline
\end{tabular}
\end{center}
    \caption{Watermarking for \textbf{different architectures} on Tiny ImageNet.}
\label{tbl:diff_arch}
\end{table}

\item \textit{Different Learning Capacity of Models.} We present the gradient search result for different learning capacity of models on Tiny ImageNet in Table \ref{tbl:diff_capacity}. ResNet~\cite{he2016deep} family is  evaluated in this experiment. The watermark signature is 60 and $\tau=15$.

\begin{table}
\begin{center}
\small
\begin{tabular}{c|c|c|c|c}
\hline
Architecture & \shortstack{Model\\Acc} & \shortstack{Watermark\\Loss} & \shortstack{Clean\\Loss} & \shortstack{Inferred\\signature} \\
\hline
ResNet18  & 52.9 & 0.036 & 0.801 & 55.2 \checkmark \\
ResNet34  & 53.6 & 0.007 & 0.679 & 55.8 \checkmark \\
ResNet50  & 54.9 & 0.019 & 0.668 & 59.5 \checkmark \\
ResNet101  & 54.6 & 0.004 & 0.736 & 58.3 \checkmark \\
ResNet152  & 56.5 & 0.007 & 0.814 & 60.4 \checkmark \\
\hline
\end{tabular}
\end{center}
    \caption{Watermarking for \textbf{different learning capacities} on Tiny ImageNet.}
\label{tbl:diff_capacity}
\end{table}

\end{enumerate}

\section{Additional Results for Watermarking Properties Analysis}
In this section, we provide more details of our experiments on analyzing the properties of watermarking.

\noindent \textbf{Resilience to Data Augmentation.} Here, we present our settings to evaluate if our proposed watermarking method could survive under common data augmentations used in neural models' training. The following data augmentations are evaluated:

\begin{enumerate}
\setlength\itemsep{0em}

\item {{\it Random Crop}~\cite{alexnet}.} Random crop is a widely used data augmentation for most neural models' training \cite{alexnet, simonyan2014very, he2016deep, yu2018deep}. It randomly crops images into smaller resolution to reduce models' overfitting on spatial location. For Tiny ImageNet, images are randomly cropped into $64 \times 64$ with a padding of 8. For Cifar, images are randomly cropped into $32 \times 32$ with a padding of 4. All our training includes random crop for best performance.

\item {{\it Horizontal Flipping}~\cite{alexnet}}. This is also a widely used data augmentation for image classification. We include horizontal flipping for all our experiments.

\item {{\it Cut Out}~\cite{devries2017improved}.} Cutout removes random region of size $M\times M$ from images at each training iteration. We set the size $M=8$ for our experiments. 

\item {{\it Label Smoothing}~\cite{szegedy2016rethinking}.} Label smoothing is also a widely used data augmentation in many tasks~\cite{muller2019does}. It reduces the probability of ground truth label  ({\it e.g.}, 100\% cat, 0\% dog, 0\% duck) by a smoothing parameter $\alpha$ and assigns probability uniformly to other classes (90\% cat, 5\% dog, 5\% duck). The smoothing parameter $\alpha$ is set to be 0.1 in our experiments. 

\item {{\it Gaussian Noise}~\cite{cohen2019certified}.} This technique simply adds noise to the input from a Gaussian distribution $\mathcal{N}(0, \sigma^2)$ to increase models' robustness. The $\sigma^2$ is set to be 0.1 in our experiments.

\item {{\it Adversarial Training}~\cite{madry2018towards}.} Neural Networks are well known for their vulnerability to adversarial attacks.~\cite{goodfellow2014explaining, madry2018towards}, and adversarial training are believed to reduce overfitting ~\cite{madry2018towards} and mitigate privacy leakage~\cite{jia2019memguard, nasr2018machine}. We address this by training with adversarial samples generated from FGSM attack~\cite{goodfellow2014explaining}. The epsilon is set to be 0.01. 

\item {{\it Differential Privacy}~\cite{dwork2008differential}.} Differential privacy is a wide adopted privacy preserving technique in many real-world applications. By adding noise to the query results, user's sensitive information cannot be inferred via querying. In our implementation, we add random noise samples from Gaussian distribution $\mathcal{N}(0, \sigma^2)$ to the output confidence. The $\sigma^2$ is set to be 0.1.

\item {{\it Color Jitter}~\cite{alexnet}.} Color jitter randomly adjusts brightness, contrast, saturation and hue of input images. We apply high intensity color augmentation in our experiments. For each color properties, the value of adjustment is randomly sampled from [-288, 288], covering $80\%$ of the range of transformation. 
The same conclusion can be made by comparing Table \ref{tbl:diff_signature} with Table \ref{tbl:aug_cj}. We further explore on CUB-200-birds with a stronger data augmentation. Watermark signature would fail to be inferred in this case. This observation implies that if models are trained heavily with similar data augmentations as watermarking, the signature inference could be confused by nearby signatures and thus fail to recover the watermarking signatures. 

\end{enumerate}

\begin{table}
\begin{center}
\small
\begin{tabular}{c|c|c|c|c}
\hline
\shortstack{Watermark\\signature} & \shortstack{Model\\Acc} & \shortstack{Watermark\\Loss} & \shortstack{Clean\\Loss} & \shortstack{Inferred\\signature} \\
\hline
60 & 52.5 & 0.042 & 0.054 & 64.8 \checkmark \\
120 & 52.5 & 0.102 & 0.244 & 108.3 \checkmark \\
180 & 53.0 & 0.052 & 0.081 & 178.8 \checkmark \\
240 & 53.0 & 0.113 & 0.132 & 254.4 \checkmark \\
300 & 51.7 & 0.125 & 0.161 & 291.3 \checkmark \\
\hline
\end{tabular}
\end{center}
\caption{Watermarking for model trained with \textbf{color jitter} augmentation for ResNet50 on Tiny ImageNet. Loss difference between clean and watermarked samples are smaller comparing with Table \ref{tbl:diff_signature}.}
\label{tbl:aug_cj}
\end{table}

In Table~\ref{tbl:aug_resnet50_tiny_imagenet}, watermark signatures can be inferred correctly from gradient search for the aforementioned data augmentations. This shows empirically that LCT is an effective watermarking approach because it is resilient to common data augmentations in neural networks' training. 
\begin{table}
\begin{center}
\small
\begin{tabular}{c|c|c|c|c|c}
\hline
Augmentation & 60 & 120 & 180 & 240 & 300 \\
\hline
\textbf{Cut Out} & 57.5 & 122.1 & 178.6 & 240.3 & 301.8 \\
\textbf{Lable Smoothing} & 59.6 & 116.9 & 181.4 & 239.3 & 299.9 \\
\textbf{Gaussian Noise} & 58.3 & 107.8 & 187.1 & 237.2 & 309.9 \\
\textbf{Adv Training} & 57.5 & 118.8 & 183.5 & 242.9 & 298.6 \\
\textbf{Differential Privacy} & 56.0 & 117.5 & 182.8 & 240.6 & 295.9 \\
\textbf{Color Jitter} & 64.8 & 108.3 & 178.8 & 254.4 & 291.3 \\
\hline
\end{tabular}
\end{center}
\caption{Inferred signatures for models trained with \textbf{different data augmentations} for ResNet50 on Tiny ImageNet.}
\label{tbl:aug_resnet50_tiny_imagenet}
\end{table}

\begin{figure}
  \begin{center}
  \begin{subfigure}[b]{\textwidth}
      \centering
      \includegraphics[width=.6\textwidth]{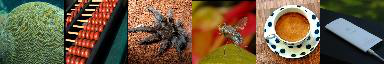}
  \end{subfigure}
  \begin{subfigure}[b]{\textwidth}
      \centering
      \includegraphics[width=.6\textwidth]{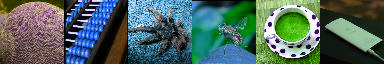}
  \end{subfigure}
  \begin{subfigure}[b]{\textwidth}
      \centering
      \includegraphics[width=.6\textwidth]{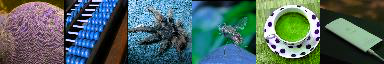}
  \end{subfigure}
  \end{center}
    \caption{Selective clean (top), watermarked (middle), reconstructed (bottom) samples' comparison for hue-based watermarking applied on the whole images.}
  \label{fig:multi_marker_samples}
\end{figure}

\noindent \textbf{Less Noticeable Watermarking.} In the previous sections, we show that LCT-based watermarking is effective against unauthorized neural learners, but may change the color property significantly in visualization, as illustrated in Figure~\ref{fig:multi_marker_samples}. One may argue that such a kind of watermarking could be too visually obvious to be recognized by unauthorized neural learners. However, given data samples, by selecting proper watermarking signature, the watermarking could be difficult to be distinguished from stylish transformation or even unnoticeable to human being. 

For example, nowadays users would often apply image filters to stylize images before publishing on social media, where filters are quite natural such as tuning the color of tree leaves from green to yellow, changing the color of sky from light blue to dark blue. The parameters of these images' filters could be used as signature for watermarking. As stylized filters are widely used, it would be difficult for neural learners to distinguish whether it is watermarking or users' preference.

\begin{figure}
  \begin{center}
  \begin{subfigure}[b]{\textwidth}
      \centering
      \includegraphics[width=.6\textwidth]{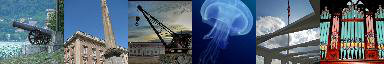}
      \caption{Clean.}
  \end{subfigure}
  \begin{subfigure}[b]{\textwidth}
      \centering
      \includegraphics[width=.6\textwidth]{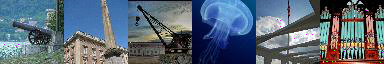}
      \includegraphics[width=.6\textwidth]{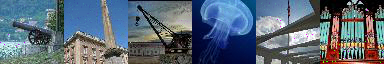}
      \caption{Watermarked (0.1) and Reconstructed (0.12).}
  \end{subfigure}
  \begin{subfigure}[b]{\textwidth}
      \centering
      \includegraphics[width=.6\textwidth]{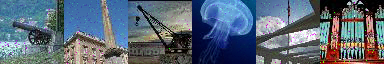}
      \includegraphics[width=.6\textwidth]{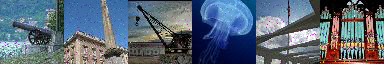}
      \caption{Watermarked (0.3) and Reconstructed (0.28).}
  \end{subfigure}
    \begin{subfigure}[b]{\textwidth}
      \centering
      \includegraphics[width=.6\textwidth]{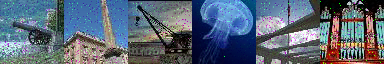}
      \includegraphics[width=.6\textwidth]{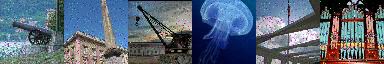}
      \caption{Watermarked (0.5) and Reconstructed (0.44).}
  \end{subfigure}
  \end{center}
    \caption{Selective samples' comparison for color-based watermarking for \textbf{partial pixels} on blue color channel. }
  \label{fig:different_alpha}
\end{figure}

Color-based transformation can be less noticeable when it is only applied on selective pixels and color channels. In particular, by following \cite{yu2001digital}, we first generate a random binary string $w$ with a fixed length $T$, and then generate pseudo-random pixels' positions as $\rho_t=(i_t, j_t)$ for each element $w_t, (1\leq t\leq T)$. Finally we change blue color channel for these pixels as:

\begin{equation}
    B_{\rho_t} \leftarrow (2w_t-1)\alpha L_{\rho_t},
\end{equation}
where $\alpha$ is the hyper-parameter of watermarking intensity and $L_{\rho_t}$ is luminance of pixel calculated by $L_{\rho_t}=0.299R_{\rho_t}+0.587G_{\rho_t}+0.114B_{\rho_t}$.

Different from~\cite{yu2001digital}, we use $\alpha$ as the watermark key and pass the pseudo-random pixels' locations and binary string $w$ to the verifier for key inference. Empirically, this kind of watermarking can be memorized by neural learners but less noticeable to human. We present visual comparison of samples in Figure~\ref{fig:different_alpha} with different value of $\alpha$. In summary, color-based watermarking can be both effective and unnoticeable with carefully selected watermarking approach and watermark keys w.r.t users' images.




In our paper, we mainly explore LCT as an effective way for anti-neuron watermarking. And we also show recovering watermark pattern fails to work in PIP while other geometrical watermarking can be applicable. There are certainly other watermarking functions that can verify unauthorized neural model training. 

\section{More Discussions on User-specific Watermarking}
As we show in previous section, signatures watermarked by user-specific LCTs would be better memorized among multiple users than those watermarked by a single LCT. This observation, on one hand, implies that selecting an arbitrary LCT for watermarking would be a good practice. It also explains why the LCT watermarking could be resilient to Color jitter. As the color transformations are well-defined and constant matrices (the matrices are fixed, the adjustment could be any value), randomly chosen users' watermarking are highly unlikely to be the same as Color jitter and would be easier to survive in data augmentation. On the other hand, it is very unlikely to infer user signatures without knowing the watermarking function as inferring from arbitrary LCTs have similar matching rate as random guessing, which makes it difficult for attackers to find out the users' signatures without knowing their watermarking functions in advance. 

\section{Comparison with Related Methods}
\noindent \textbf{Comparing to Membership Inference Attacks.} As membership inference attacks (MIAs) determine membership of given data, it is reasonable to consider such methods in personal data protection. However, there are three restrictions preventing MIAs from being applied in real PIP scenarios. 

\textbf{(i)} The first restriction is MIAs require prior knowledge of training data, which would be difficult to obtain as a common user. For example, user needs auxiliary data which has the same data distribution and similar data size as training data so that the well-trained shadow models could have similar performance as adversary models~\cite{shokri2017membership}. Or users need to know some samples that are in adversary model's training data~\cite{yeom2018privacy}. 

\textbf{(ii)} If users exploit their own data and train shadow models to perform MIAs, the success accuracy would still be low. This is because adversary could train a much better models by using much more data than a single user. In this case, the training loss of adversary model would be much less than the users shadow models. As we discussed in the main draft, such performance difference would lead to poor classification accuracy in membership inference.

On the other hand, as neural learner collects data from many others, there are chances that users' data distribution are slightly different ({\it i.e.}, users watermarking their data in different ways or applying different filters). Under this circumstance, MIAs would possibly perform worse as a single threshold might not work well for heterogeneous users data.

\textbf{(iii)} Last but not the least, even if we assume the verifier obtains such knowledge, the inference results would be unconvincing for an arbitration. As MIAs only produce binary outputs ({\it i.e.}, True of False), the probability of a correct guess is already $50\%$. Further, MIAs determine membership by considering data in training must have a small loss. But a small loss of user data can not guarantee user data were used for training. Such a result could possibly be caused by learning similar data rather than the user's. 

In summary, MIAs would be less practical than watermarking in personal data protection.

Early studies on MIAs consider watermarking as a special case. In the study~\cite{yeom2018privacy} on membership inference, they discuss the membership advantage~\cite{yeom2017unintended} is not necessary if there exists features as a prior knowledge that can be used to distinguish data, ({\it e.g.} unique id for each image). This follows same idea of watermarking and verification discussed in the paper. However, \cite{yeom2018privacy}'s study does not study what watermarking technique can be used to against neural modeling training. They assume the all users can substitute their original data with an identifier via an arbitrary function $G$~\cite{garay2014advances} and this substitution would not interfere the embedding identifiers. This might work if neural networks can perfectly memorize every detail of training data, including the identifier. However, as shown in our experiments, this is not true. Different users with the same watermarking function would interfere each other heavily, as neural models can learn the data augmentation during training~\cite{zhang2021understanding}. One contribution of this paper is to show that LCT is an effective anti-neuron watermarking method against neural model training in various realistic settings.

To justify the above discussions, we first propose a new evaluation metric for protection, which quantifies both inference accuracy and signature space, as discussed in the main manuscript. Then we conduct experiments on two PIP cases where users train their own shadow models and different users exploit different watermarking. To perform MIAs, we assume the verifier obtains the necessary knowledge of training distribution, and compare our watermarking method with two state-of-art MIAs: MIA-std~\cite{yeom2018privacy} and MIA-pow~\cite{sablayrolles2019white}. The first MIA exploits few samples known from model's training and use the average training loss as a threshold to determine the membership. If the loss of given testing sample is smaller than the threshold, the sample would be used for training. The second method requires extra samples known from the held-out set and find the best threshold for both training and held-out samples.

For the experiment, we follow similar settings in~\cite{kaya2021does} for these two attacks. To compare with watermarking fairly, we conduct MIA on average loss of user data, not data per se, which would increase accuracy of MIAs. In the case when user exploit their own data to perform MIAs, we assume 10 users and each user utilize his/her data to train shadow models and use the knowledge of shadow models to further perform MIAs, respectively; We use MIA-std~cite{yeom2018privacy} as the baseline method and obtain the decision threshold using average training loss from user's shadow models. In the case when users' data distribution are slight different, we randomly split Tiny ImageNet into 1,000 users for training set and 100 users for validation set, with 100 data samples for each user. For MIA-std~\cite{yeom2018privacy}, we randomly select 5 users from training and calculate the average loss as the membership threshold. For MIA-pow~\cite{sablayrolles2019white}, we randomly select 5 users from training and 5 user from validation to search for the best threshold. Then we test the inference accuracy on 100 user data samples, with 50 randomly chosen from training and 50 randomly chosen on validation. Note that the testing users would be exclusive from users used for finding threshold. 

\begin{figure}
\begin{center}
    \includegraphics[width=.6\textwidth]{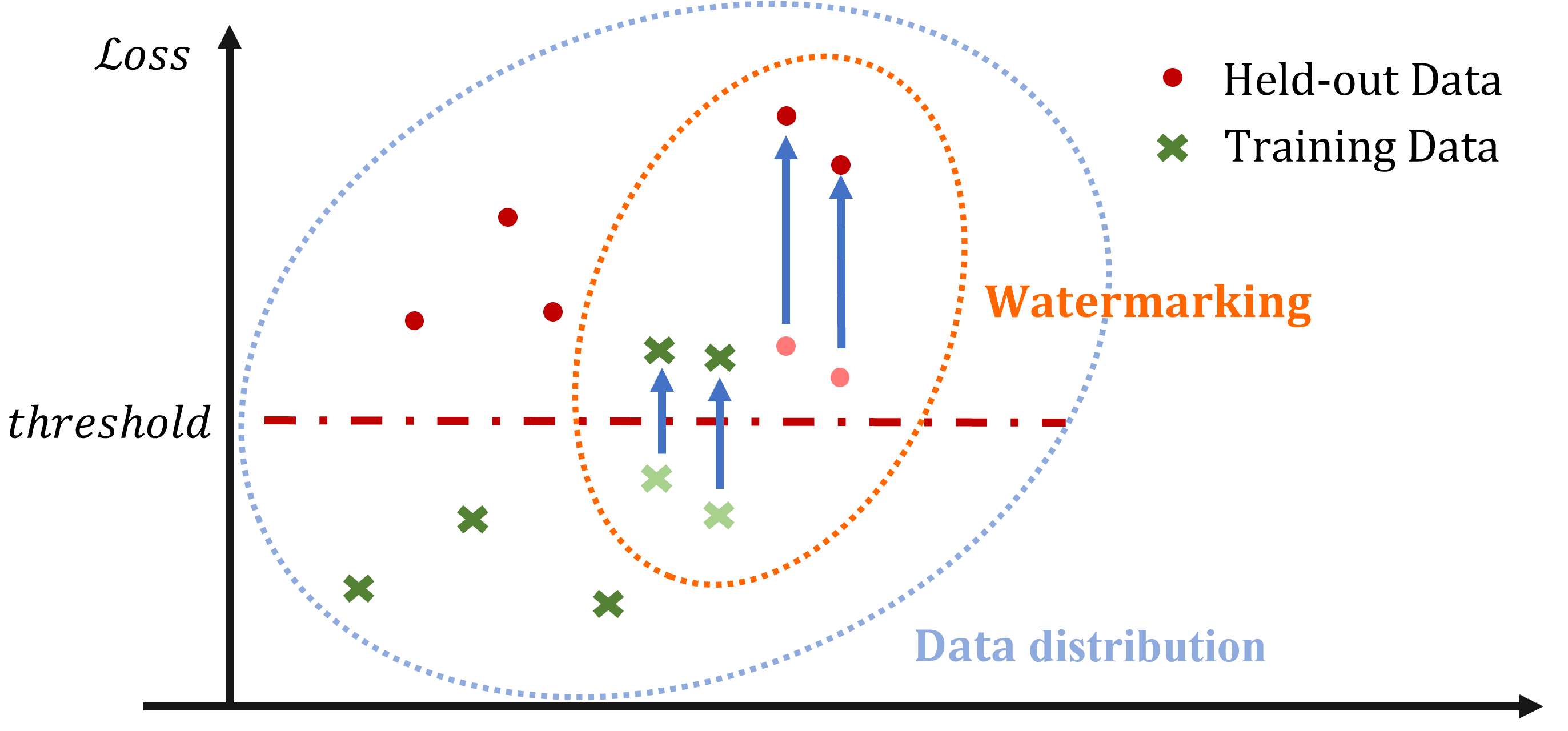}
\end{center}
\caption{\textbf{Illustration of why MIAs would decrease performance when partial users exploit watermarking.} As watermarking function lifts loss of user data, the original MIA threshold calculated by unwatermarked data would misclassify the membership of watermarked data and thus degrades performance. The experiment result for memorization in the main draft verifies the above idea.}
\label{fig:lift}
\end{figure}

From Table~\ref{tbl:mia_acc}, we show the matching accuracy for MIAs and our method on these testing users. It can be observed that the performance of MIAs will first decrease and then increase as ratio grows. Specifically, when fewer user data are watermarked, the MIAs' threshold would be mainly determined by the unwatermarked users data. Meanwhile, as some user data exploit watermarking, the loss of watermarked data would be lifted, as illustrated in Figure~\ref{fig:lift}, and even if the watermarked samples were used in training, the MIAs would misclassify these samples. The situation would be worst when there are similar amount of watermarking data and unwatermarking data. Such a result show that MIAs would be unstable when user data are heterogeneous. 

\begin{table}
\begin{center}
\small
\begin{tabular}{l|c|c|c|c|c|c}
\hline
Ratio & 0\% & 20\% & 40\% & 60\% & 80\% & 100\% \\
\hline
MIA-std & 77.8 & 74.6 & 67.3 & 72.2 & 69.5 & 79.0 \\
MIA-pow & 92.7 & 84.8 & 77.6 & 80.3 & 76.5 & 94.0 \\
\hline
ANW & - & 96.0 & 95.2 & 92.6 & 86.5 & 82.0 \\
\hline
\end{tabular}
\end{center}
\caption{\textbf{The matching accuracy between membership inference attacks and anti-neuron watermarking}. Different ratio of users exploit different LCT for their data. In MIAs, a match would be calculated by binary inference result, but in ANW, a match would imply the correct signature, which would be more difficult because there are 12 signature values in this experiment. }
\label{tbl:mia_acc}
\end{table}

\noindent \textbf{Comparing to Dataset Tracing.} Recent studies~\cite{sablayrolles2020radioactive, li2020open, zhang2020model} on dataset protection investigate watermarking against neural training and it is tempting to apply these methods in personal data protection. However, as we discussed in the main manuscript, dataset tracing~\cite{sablayrolles2020radioactive, li2020open, zhang2020model} requires full knowledge of training data, so that they can generate ``watermarked data'' by exploiting pretrained classifier on the dataset as adversarial training~\cite{madry2018towards}. However, from users' perspective, such prior knowledge would not be sufficient because it is impossible for a common user to know how other data will be collected and what tasks will be performed. As a result, techniques like dataset tracing~\cite{sablayrolles2020radioactive, li2020open, zhang2020model} can not be directly applied to the PIP scenario. 

As dataset tracing~\cite{sablayrolles2020radioactive} can be applied on partial data, we compare our method with dataset tracing~\cite{sablayrolles2020radioactive} and verify whether both methods can work well with very limited watermarking data, which is common in the PIP problem. To make dataset tracing work, we pre-train a ResNet50 model on Tiny ImageNet and adopt the same setting as ~\cite{sablayrolles2020radioactive} to generate ``radioactive data''. As shown in the main manuscript, with $0.1\%$ data being watermarked, our LCT method is more effective as we observe that watermarking signature can be memorized well as expected. 

\section{Memorization Analysis of Watermarking}

\begin{figure}
\begin{center}
    \includegraphics[width=200pt,height=144pt]{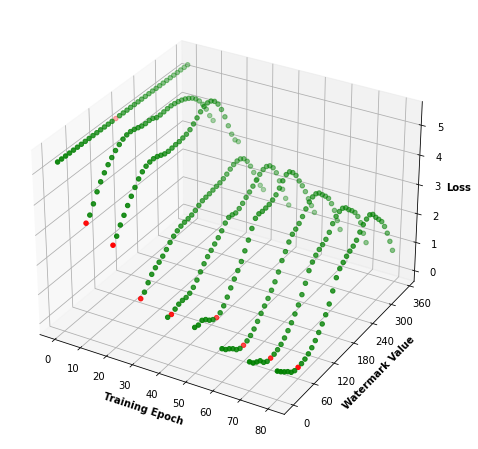}
\end{center}
\caption{The watermarking signature can be memorized before the end of training. The \textbf{red point} indicates inferred signature that achieves minimal loss over other signatures.}
\label{fig:training_process}
\end{figure}

\noindent \textbf{When Signature is being Memorized during Neural Model Training?} One interesting problem for watermarking is when a signature is memorized by neural models. From previous studies on MIAs~\cite{shokri2017membership, yeom2018privacy, kaya2021does, Song2020Overlearning}, user privacy information is being leak when model's overfitting. As a result, it is likely for models to memorize watermarking signatures when models over-learn the watermarked data~\cite{Song2020Overlearning}. To explore the answer of this question, we infer signatures by grid search for different checkpoints of model during training. In Figure~\ref{fig:training_process}, we can observe the inferred signature reaches 0 at early stage of training and gradually reach 60 with more learning epochs. This result illustrates that the watermarking signature could be memorized before the end of training. This empirical results can also explain why data augmentation (watermarking) can be learned by neural models during training, according to the study~\cite{zhang2021understanding} for neural generalization.

\noindent \textbf{Watermarking Increase Memorization of User Data.} In the main draft, we mention how to use ``memorization value estimate''~\cite{feldman2020neural} to study how watermarking improve memorization of user data. As user data is watermarked, the probability density of user data would be lifted into low density region and thus being easier memorized by neural learners. During training process, the watermarking signature would be better memorized along with user data. According to the study~\cite{feldman2020neural} on memorization, for training algorithm $A$ on a dataset $S=((x_1, y_1),...,(x_n,y_n))$, the amount of label memorization by $\mathcal{A}$ on example $(x_i, y_i)\in S$ is defined as,

\begin{equation}
    mem(\mathcal{A}, S, i):=\underset{h \leftarrow \mathcal{A}(S)}{\mathbb{P}_r}[h(x_i)=y_i]
    -\underset{h \leftarrow \mathcal{A}(S^{\backslash i})}{\mathbb{P}_r}[h(x_i)=y_i].
\end{equation}

To estimate above memorization value on sample index at $i$, \cite{feldman2020neural} firstly selects random subsets for $S$, with some subsets including sample $i$ and some exclude sample $i$. Then, $K$ models are being trained using these subsets, grouped into 2, one includes sample $i$ and the other excludes sample $i$. The memorization value estimate (MAE) is finally calculate by averaging the difference of $\mathbb{P}_r$ between these two groups of models. In our experiments, we split the training set of Tiny ImageNet into 1,000 users and choose one split as user data (we calculate MAE on a collection user data instead of one sample). Then we randomly use $70\%$ of users data to construct 20 subsets. 20 models are being trained correspondingly to calculate the final MAE. We fix the indices for this experiments and train 20 models with and without watermarking on user data. From the results we show in the main manuscript, the MAE of user data increase after watermarking, indicating that watermarking improves model memorization ability on the given user data.

\end{document}